\documentclass[twocolumn,showpacs,preprintnumbers,amsmath,amssymb]{revtex4}

\usepackage{graphicx}
\usepackage{dcolumn}
\usepackage{bm}
\usepackage{amssymb}
\usepackage{amsbsy}
\usepackage{amsmath}
\usepackage{mathrsfs}

\newcommand{\be}{\begin{equation}}
\newcommand{\ee}{\end{equation}}

\begin{document}

\title{Studying Soft Matter with ``Soft'' Potentials: Fast Lattice Monte Carlo Simulations and Corresponding Lattice Self-Consistent Field Calculations}

\author{Qiang Wang}
\email{q.wang@colostate.edu}
\affiliation{Department of Chemical and Biological Engineering, Colorado State University, Fort Collins, CO 80523-1370}

\date{\today}

\begin{abstract}
The basic idea of fast Monte Carlo (MC) simulations is to perform particle-based MC simulations with the excluded-volume interactions modeled by ``soft'' repulsive potentials that allow particle overlapping. This gives much faster system relaxation and better sampling of the configurational space than conventional molecular simulations with ``hard'' repulsions that prevent particle overlapping. Here we present fast lattice MC (FLMC) simulations for confined homopolymers, where multiple occupancy of lattice sites is allowed with a proper Boltzmann weight and thus the evaluation of nearest-neighbor interactions can also be avoided. When compared with the corresponding lattice field theories based on the \emph{same} Hamiltonian, FLMC simulations further provide a powerful means for unambiguously and quantitatively revealing the correlation/fluctuation effects.
\end{abstract}

\maketitle

Significant advances have been made in the past three decades in applying conventional, ``particle-based'' molecular simulation techniques (e.g., Monte Carlo (MC) and molecular dynamics simulations) to the study of polymeric systems. Even with coarse-grained models where each segment represents several to tens of repeat units of real polymers, however, conventional molecular simulations of multi-chain systems are still hindered by ``hard'' excluded-volume interactions and, for off-lattice simulations, expensive pair-potential calculations. The former is implemented in off-lattice simulations commonly by either hard-sphere repulsion or the harsh Lennard-Jones repulsion, and in lattice simulations by the self- and mutual-avoiding walk (SMAW). While such hard-core repulsions are needed to produce realistic dynamics, they greatly reduce the chain relaxation towards equilibrium configurations as well as the efficiency of sampling the configurational space. On the other hand, the pair-potential calculations are expensive for dense polymeric systems, and lattice simulations are in general much faster and thus widely used.

The basic idea of fast MC simulations is to perform particle-based MC simulations with the excluded-volume interactions modeled by ``soft'' repulsive potentials that allow particle overlapping (i.e., by soft particles whose interaction energy $u(r)$ is finite when they overlap). This avoids the hard repulsions (i.e., $u(r \to 0) \to \infty$ as in the Lennard-Jones potential) used in conventional molecular simulations, thus allowing much faster chain relaxation and better sampling of the configurational space. Such coarse-grained models are particularly suitable for the study of equilibrium properties of soft matter such as polymers. It turns out that soft potentials are commonly used in polymer field theories (e.g., the widely applied self-consistent field (SCF) theory with great success for many polymeric systems)\cite{GlenBK}, where individual polymer segments are modeled as volumeless points with the excluded-volume interactions described by either the Helfand compressibility\cite{Helf} or the incompressibility constraint. Using the \emph{same} Hamiltonian in both fast MC simulations and polymer field theories therefore enables quantitative comparisons between them \emph{without any parameter-fitting} to unambiguously reveal the consequences of approximations in the theories (e.g., the effects of correlations and fluctuations neglected in SCF theory)\cite{FRI}. Soft potentials have also been used in dissipative particle dynamics\cite{DPD} and the dynamic mean-field density functional method\cite{DDFT} that focus on the system dynamics.

In our recent work\cite{WQ-FOMC}, fast off-lattice MC (FOMC) simulations were performed for several systems, using an isotropic and position-independent soft repulsive potential. As studied in detail there, while a spatial discretization scheme used in previous FOMC simulations\cite{FOMC} avoids pair-potential calculations, it is equivalent to an \emph{anisotropic and position-dependent} pair potential that cannot be implemented in field theories. In this Letter, we report fast lattice MC (FLMC) simulations where multiple occupancy of lattice sites is allowed with a proper Boltzmann weight. While this Domb-Joyce model\cite{Domb} was already studied in the early single-chain simulations of variable excluded-volume effects\cite{KRA}, we demonstrate here its great advantages for multi-chain systems and further compare our FLMC results with the corresponding lattice self-consistent field (LSCF) calculations based on the \emph{same} Hamiltonian.

As a simple example, we consider an inhomogeneous system of $n$ homopolymer chains each consisting of $N=60$ segments confined between two parallel and impenetrable walls placed in the $x$ direction. The energetic wall-polymer interaction is ignored, and the system Hamiltonian due to non-bonded interactions is given by
\be \label{eq:HE1}
\beta \mathcal{H}^E = \frac{1}{2 \kappa \rho_0} \sum_\mathbf{r} [\hat{\rho}(\mathbf{r}) - \rho_0]^2
\ee
where $\beta \equiv 1 / k_B T$ with $k_B$ being the Boltzmann constant and $T$ the absolute temperature, the average segmental density $\rho_0 \equiv nN/V$ with $V$ being the total number of lattice sites, the microscopic density of polymer segments at a lattice site ${\bf r}$ is $\hat{\rho}({\bf r}) \equiv \sum_{k=1}^{n} \sum_{s=1}^N \delta_{{\bf r},\mathbf{R}_{k,s}}$ with ${\bf R}_{k,s}$ denoting the lattice position of the $s^{\rm th}$ segment on the $k^{\rm th}$ chain and $\delta_{{\bf r},\mathbf{R}_{k,s}}$ being the Kronecker $\delta$-function, and $\kappa = 0.48$ denotes the compressibility. The value of $N/\kappa$ indicates that our system is nearly incompressible. Note that Eq.~(\ref{eq:HE1}) can equivalently represent homopolymers in an implicit, good solvent\cite{WQ-FOMC,Alfr1}. Hereafter we define two physical parameters that determine the system behavior: chain number density $C \equiv n/V$ and excluded-volume parameter $B \equiv N/\kappa C$.

Our FLMC simulations are performed in a canonical ensemble on a simple cubic lattice of $V = L_x L^2$ sites, where $L_x=10$ is the number of lattice sites in the $x$ direction that can be occupied by polymers, and $L = 20\sim 60$ (depending on the finite-size effects) is the number of lattice sites in both the $y$ and $z$ directions along which the periodic boundary conditions are applied. Our trial moves include local moves (end-rotation and kink-jump), reptation, and bond-rotation (where a randomly selected bond vector of a chain is altered), occurring with probabilities of 0.1, 0.45, and 0.45, respectively. About $2 \times 10^8 \sim 1.3 \times 10^{10}$ trial moves are performed in each simulation (depending on $n$ and $V$), and the Metropolis acceptance criterion is used. We have estimated the error bar of each ensemble-averaged quantity in FLMC simulations as three times its standard deviation, with the statistical correlations among samples collected after equilibration taken into account through the block analysis\cite{Allen}; the error bar is smaller than the symbol size in all the cases and thus not shown.

To formulate the corresponding LSCF theory, we start from the canonical-ensemble partition function
\be \label{eq:z0}
\mathcal{Z} = \frac{1}{n!} \sum_{\{{\bf R}_{k,s}\}} \exp\{- \beta \mathcal{H}^C[\{{\bf R}_k\}] - \beta \mathcal{H}^E\}
\ee
where the summation is over all possible lattice positions of all polymer segments, and $\mathcal{H}^C$ is the Hamiltonian due to chain connectivity; $\beta \mathcal{H}^C = 0$ if the chain connectivity is maintained for all chains on the lattice, and $\infty$ otherwise. After the Hubbard-Stratonovich transformation, i.e., inserting in Eq.~(\ref{eq:z0}) the identity $1 = \int \mathscr{D}\phi \mathscr{D}\omega \exp \{ \sum_\mathbf{r} \omega(\mathbf{r}) [ \rho_0 \phi(\mathbf{r}) - \hat{\rho}(\mathbf{r}) ] \}$ where $\phi(\mathbf{r})$ is the (normalized) segmental density field constrained to $\hat{\rho}(\mathbf{r})/\rho_0$ and $\omega(\mathbf{r})$ is the (purely imaginary) conjugate field imposing the constraint, we finally have $\mathcal{Z} = \int \mathscr{D}\phi \mathscr{D}\omega \exp\{- n \beta f_c[\phi,\omega]\}$ with
\be \label{eq:fc}
\beta f_c = \frac{N}{2 \kappa V} \sum_{\bf r} [\phi({\bf r}) - 1]^2 - \frac{1}{V} \sum_{\bf r} \omega({\bf r}) \phi({\bf r}) - \ln Q[\omega]
\ee
and the single-chain partition function $Q \equiv \sum_{\{{\bf R}_s\}} \exp[-\beta \mathcal{H}^C/n - \sum_{s=1}^N \omega({\bf R}_s)/N] / \sum_{\{{\bf R}_s\}} \exp(-\beta \mathcal{H}^C/n)$, where we have omitted a constant factor in $\mathcal{Z}$ and re-scaled variables according to $N \omega(\mathbf{r}) \to \omega(\mathbf{r})$. The SCF equations are obtained by setting $\delta \beta f_c / \delta \phi({\bf r}) = \delta \beta f_c / \delta \omega({\bf r}) = 0$ (i.e., the mean-field approximation) and given by
\be \label{eq:omega}
\omega({\bf r}) = (N/\kappa) [\phi({\bf r}) - 1]
\ee
\be
\phi(\mathbf{r}) = \frac{\exp \left[\omega({\bf r})/N\right]}{N Q} \sum_{s=1}^N q_s(\mathbf{r}) q_{N+1-s}(\mathbf{r})
\ee
where $q_s(\mathbf{r})$ is the probability of finding a partial chain of $s$ segments starting anywhere in the system and ending at ${\bf r}$. According to the chain connectivity, we have
\be \label{eq:CKE}
q_{s+1}({\bf r}) = \frac{\exp \left[-\omega({\bf r})/N\right]}{z} \sum_{{\bf r}_n} q_s({\bf r}_n), \quad q_0({\bf r})=1
\ee
where ${\bf r}_n$ denotes a nearest neighbor of ${\bf r}$ on the lattice and $z=6$ is the lattice coordination number. Finally,
\be
Q = \frac{1}{V} \sum_\mathbf{r} \exp\left[\frac{\omega({\bf r})}{N}\right] q_s(\mathbf{r}) q_{N+1-s}(\mathbf{r})
\ee

Note that our LSCF formalism is different from that of Scheutjens and Fleer\cite{SF}: In addition to the incompressibility constraint, their model uses nearest-neighbor interactions between different species. Since multiple occupancy of lattice sites is allowed in our model, however, we can use $\delta_{{\bf r},{\bf r}'}$ instead (refer to Eq.~(\ref{eq:HE1})). This further avoids the evaluation of nearest-neighbor interactions in our FLMC simulation, which makes it very fast.

For the confined homopolymers, we solve the SCF equations in 1D using the Broyden method combined with a globally convergent strategy\cite{Broy}, where the residual errors of Eq.~(\ref{eq:omega}) at all $x$ are less than $10^{-12}$. We then calculate the propagator $Q_{p,s}(x|x')$, which corresponds to the probability of finding a partial chain of $s+1$ segments that starts at $x'$ and ends at $x$ in the obtained conjugate field $\omega(x)$, from the following equation analogous to Eq.~(\ref{eq:CKE})
{\setlength \arraycolsep{1pt}
\begin{eqnarray}
Q_{p,s+1}(x|x') = & & \exp[-\omega(x)/N] \{z_0 Q_{p,s}(x|x') \nonumber \\
& & + z_1 [Q_{p,s}(x+1|x') + Q_{p,s}(x-1|x')] \} \nonumber
\end{eqnarray} }
with the initial condition of $Q_{p,s=0}(x|x') = \delta_{x,x'}$ and the appropriate boundary conditions, where $z_0=2/3$ is the fraction of nearest-neighbor lattice sites at the same $x$ and $z_1=(1-z_0)/2$. The mean-square chain radius of gyration in the $x$ direction can finally be computed as
\[
\langle R_{g,x}^2 \rangle = \frac{1}{N^2} \sum_{s=1}^{N-1} (N-s) \frac{\sum_x \sum_{x'} Q_{p,s}(x|x') (x-x')^2}{\sum_x \sum_{x'} Q_{p,s}(x|x')}
\]
where the last term in the summation (i.e., with $s=N-1$) is the mean-square chain end-to-end distance in the $x$ direction.

\begin{figure}[t]
\begin{center}
\includegraphics[height=5.5cm]{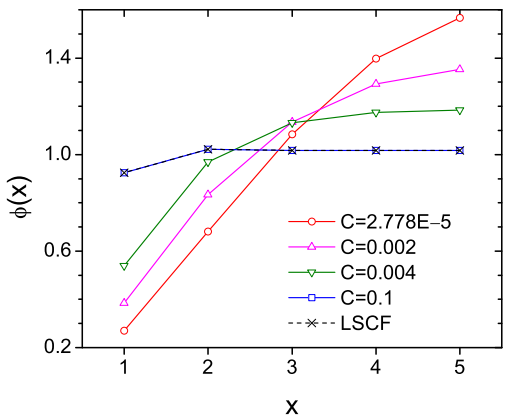}
\end{center}
\caption{\label{phi}(Color online) Polymer segmental density profiles $\phi(x)$ from FLMC simulations at different chain number densities $C$, and from LSCF calculations (which correspond to $C \to \infty$). $x=1$ is the closest lattice layer to a confining wall, and $x=5$ is in the middle of the film.}
\end{figure}

Fig.~\ref{phi} shows the ensemble-averaged segmental density profiles $\phi(x) \equiv \langle \hat{\rho}(x) \rangle / \rho_0$ from FLMC simulations at different chain number densities $C$; since $\phi(x)$ is symmetric about the mid-plane between the two confining walls, it is shown only from $x=1$ (closest to a wall) to 5 (in the middle of the film). The smallest $C$ value in FLMC simulations is obtained from a single chain with $L=N$, which is effectively self-avoiding due to the large corresponding $B$ value. As $C$ increases, $\phi(x)$ increases near the walls and decreases in the interior of the film. This is in good agreement with recent field-theoretic simulations\cite{Alfr1}. $\phi(x)$ from LSCF calculations is also shown in Fig.~\ref{phi}, which corresponds to $C \to \infty$ and is approached by FLMC simulations with increasing $C$. For $C \gtrsim 0.1$, $\phi(x)$ is oscillatory (i.e, exhibits local maxima at $x=2$ and 4) rather than monotonic with $x$ due to the packing effects.

\begin{figure}[t]
\begin{center}
\includegraphics[height=5.5cm]{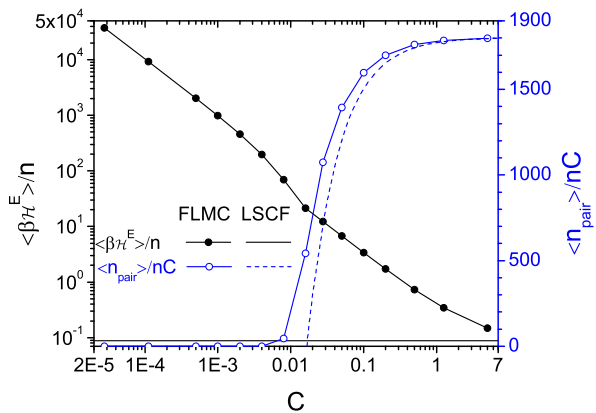}
\end{center}
\caption{\label{hE}(Color online) Ensemble-averaged non-bonded interaction energy per chain $\langle \beta \mathcal{H}^E \rangle / n$ and number of overlapping segment pairs occupying the same lattice sites $n_{\rm pair}$.}
\end{figure}

Fig.~\ref{hE} shows how the ensemble-averaged non-bonded interaction energy per chain $\langle \beta \mathcal{H}^E \rangle / n$ varies with $C$. Its behavior can be well understood by considering the number of overlapping segment pairs occupying the same lattice sites $n_{\rm pair}$ also shown in the figure; note that $\beta \mathcal{H}^E / n = (N / 2\kappa) [(2 / N^2) (n_{\rm pair} / n C) + 1 / N C - 1]$. At small $C$ values ($\lesssim 0.002$) we effectively have SMAW (i.e., $\langle n_{\rm pair} \rangle = 0$) due to the large corresponding $B$ values, while at large $C$ values $\langle n_{\rm pair} \rangle / n C$ approaches that predicted by LSCF theory, which is obtained by equating $\beta \mathcal{H}^E / n$ to the first term on the right-hand-side of Eq.~(\ref{eq:fc}) after LSCF equations are solved. Due to its mean-field approximation, LSCF theory underestimates $n_{\rm pair}/nC$, and gives $n_{\rm pair} < 0$ for $C<1.664 \times 10^{-2}$.

\begin{figure}[t]
\begin{center}
\includegraphics[height=5.5cm]{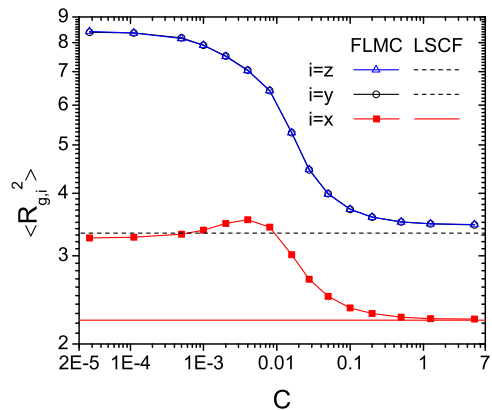}
\end{center}
\caption{\label{Rg2}(Color online) Log-log plot of mean-square chain radii of gyration $\langle R_{g,i}^2 \rangle$ along different directions $i$ as a function of chain number density $C$.}
\end{figure}

Fig.~\ref{Rg2} shows how the mean-square chain radii of gyration $\langle R_{g,i}^2 \rangle$ ($i=x,y,z$) change from the (self-avoiding) single-chain case to the LSCF limit as $C$ increases. In the directions parallel to the confining walls (i.e., $i=y$ and $z$), $\langle R_{g,i}^2 \rangle$ monotonically decreases with increasing $C$ and approaches the LSCF limit corresponding to the random walk where $\langle R_{g,i}^2 \rangle = (N^2-1) / 18 N$. In the direction perpendicular to the walls, however, $\langle R_{g,x}^2 \rangle$ first increases with increasing $C$ to a maximum located around $C=0.004$, then decreases to approach the LSCF limit. The same behavior is also found in the mean-square chain end-to-end distances (data not shown).

To summarize, we have presented for the first time FLMC simulation data ranging from the single-chain case all the way to the LSCF limit. In terms of the invariant degree of polymerization $\bar{\mathcal{N}} \equiv [n / (V/R_{e,0}^3)]^2$, where the root-mean-square end-to-end distance of ideal chains is given by $R_{e,0} = \sqrt{N-1}$ for our lattice polymers, these data correspond to $\bar{\mathcal{N}} \approx 1.585 \times 10^{-4} \sim 5.134 \times 10^6$, more than ten orders of magnitude across. Even better, the total simulation time for obtaining these data is less than a couple of days on an Intel Core 2 Q6600 2.4 GHz processor. This clearly demonstrates the high efficiency of FLMC simulations (with multiple occupancy of lattice sites and Kronecker $\delta$-function interactions) over both conventional lattice MC simulations (with SMAW and nearest-neighbor interactions) and FOMC simulations (with pair-potential calculations). When compared with the corresponding lattice field theories based on the \emph{same} Hamiltonian, FLMC simulations further provide a powerful means for unambiguously and quantitatively revealing the correlation/fluctuation effects in the system. Although here we use confined compressible homopolymers as a simple example, FLMC simulations can be readily performed for more complex systems such as polymer blends and block copolymers or even incompressible systems (where $\hat{\rho}({\bf r}) = \rho_0$ is required at all lattice sites but not SMAW). These will be the topics of our future publications.

Financial support for this work was provided by NSF CAREER Award CBET-0847016.

\bibliographystyle{unsrt}

\end{document}